\documentclass[prd,twocolumn,nopacs,nofootinbib]{revtex4}

\usepackage{amsfonts}
\usepackage{amsmath}
\usepackage{amssymb}
\usepackage{bm}
\usepackage{tipa}

\newcommand{\be}{\begin{equation}}
\newcommand{\ee}{\end{equation}}
\newcommand{\ba}{\begin{eqnarray}}
\newcommand{\ea}{\end{eqnarray}}

\newcommand{\bbe}{\boldsymbol{\mathrm{e}}}
\newcommand{\e}{\mathrm{e}}
\newcommand{\ie}{\text{\textschwa}}

\newcommand{\bA}{\boldsymbol{A}}
\newcommand{\bF}{\boldsymbol{F}}

\newcommand{\bT}{\boldsymbol{T}}

\newcommand{\diff}{\textrm{d}}
\newcommand{\Diff}{\textrm{D}}

\newcommand{\lp}{\left(}
\newcommand{\rp}{\right)}
\newcommand{\lb}{\left[}
\newcommand{\rb}{\right]}

\newcommand{\nn}{\nonumber}

\begin{document}

\title{Scale-invariant cosmology in de Sitter gauge theory}

\author{Tomi S. Koivisto}
\email{tomik@astro.uio.no}
\address{Laboratory of Theoretical Physics, Institute of Physics, University of Tartu, W. Ostwaldi 1, 50411 Tartu, Estonia}
\address{National Institute of Chemical Physics and Biophysics, R\"avala pst. 10, 10143 Tallinn, Estonia}
\author{Luxi Zheng}
\email{luxi.zheng@ut.ee}
\address{Laboratory of Theoretical Physics, Institute of Physics, University of Tartu, W. Ostwaldi 1, 50411 Tartu, Estonia}

\begin{abstract}

The Planck mass and the cosmological constant determine the minimum and the maximum distances in the physical universe.
A relativistic theory that takes into account a fundamental distance limit $\ell$ {\it on par} with the fundamental speed limit $c$, 
is based on the de Sitter extension of the Lorentz symmetry. This article proposes a new de Sitter gauge theory of 
gravity which allows the consistent cosmological evolution of the $\ell$. The theory is locally equivalent to Dirac's scale-invariant
version of general relativity, and suggests a novel non-singular extension of cosmology.   

 
\end{abstract}

\maketitle


\section{Introduction}

In the standard $\Lambda$CDM model of cosmology \cite{Aghanim:2018eyx}, the background universe is dS (de Sitter). The dS geometry can be seen as a 4-dimensional hyperboloid of curvature $R_\Lambda$ and the radius $\ell_\Lambda = \sqrt{3/\Lambda}=\sqrt{12/R_\Lambda}$ embedded in a  5-dimensional Minkowski space. The dS scale introduces a horizon, the maximum proper distance up to which any signal can reach.  

At the other end of scales, the space has a resolution limit given by the Planck length $\ell_P$, since the wavelengths of photons required to probe smaller distances would have enclosed the photon's energy within its Schwarzschild radius. There are more refined thought experiments that lead to the existence of a minimum length, and it is either assumed or predicted in most of the approaches to quantum gravity \cite{Garay:1994en}. 

An observer-independent scale $\ell$ is naturally incorporated into the physical theory of the universe by postulating the spacetime symmetry SO(4,1) instead of the usual ISO(3,1). Analogously to the Galilean group being the $c \rightarrow \infty$ contraction limit of the Poincar\'e group ISO(3,1), the latter is the contraction limit $\ell \rightarrow \infty$ of the dS group SO(4,1) \cite{Dyson:1972sd}. An extension of the relativity principle that describes the kinematics with a finite limiting distance $\ell$ has been formulated as the projective special \cite{Licata:2017dpm}, the doubly special \cite{AmelinoCamelia:2002wr} and the dS special \cite{Aldrovandi:2006vr} relativity. The gravitational theory is obtained by localisation of the symmetry \cite{Westman:2014yca}.  

In this paper we propose a dS gauge theory with a dynamical dS scale $\ell(x)$. 
Since the Planck length $\ell_P$ is defined as 
\be
\ell_P = \sqrt{\frac{\hbar G}{c^3}}\,,\label{planck}
\ee
where from now on we will set the speed of light $c=1$ and the Planck constant $\hbar=1$ to unity, the dynamical Planck length $\ell=\ell(x)$ could equivalently be considered as the dynamical (squareroot of the) Newton's constant $G=G(x)$. A well-known realisation of this aspect of the theory is scalar-tensor gravity, which promotes the gravitational coupling into a dynamical scalar field \cite{Brans:1961sx}. 

Indeed, we will arrive at an action which is equivalent to the conformally coupled scalar-tensor gravity \cite{Dirac:1973gk}, and wherein the $\ell$ plays the role of a dilaton field. While the dilaton is usually introduced in the context of Weyl gauge theory \cite{Blagojevic:2002du,Blagojevic:2013xpa,Scholz:2017pfo}, the kinematical origin of scale invariance in a dS gauge theory seems not to have been clarified previously.  



It was shown long ago that dS gravity can be reduced to Einstein's gravity with a cosmological constant \cite{MacDowell:1977jt,Pagels:1983pq}, and nowadays it has been well understood that the implied symmetry-breaking is but a realisation of Cartan's original geometrical construction \cite{Wise:2006sm}. A fine introduction to the dynamical symmetry breaking in Cartan geometry, and the most general polynomial form of such a theory, were presented in \cite{Westman:2014yca}. A physical observer requires the further breaking \cite{Gielen:2012fz} of SO(4,1)$\rightarrow$SO(3,1)$\rightarrow$SO(3), where the final step could be the geometrical origin of cold dark matter \cite{Zlosnik:2018qvg}, the CDM part of the $\Lambda$CDM \cite{Aghanim:2018eyx}.      


This framework provides a robust approach also to the problems with the $\Lambda$ \cite{Pagels:1983pq}, whilst paving the way towards reconciliation of gravity and quantum mechanics by lifting the kinematics of dS special relativity \cite{Aldrovandi:2006vr} to the dynamics of dS general relativity \cite{Aldrovandi:2007bi}. 
The proper formulation of a dS gauge theory as a Cartan geometry where the homogeneous model spaces are flat and their scale $\ell$ is a function of the coordinates in the quotient dS spacetime has been already developed by Jennen and Pereira \cite{Jennen:2014mba,Jennen:2015bxa,Jennen:2016pqw}.

In this article we propose a completion and generalisation of their theory and explore its cosmological solutions. We begin in Section \ref{gauge} by reviewing the gauge theory of the dS group in Cartan geometry, now specifically adapted to cosmology. In Section \ref{cosmo} we study the cosmological background solutions and derive a family of exact solutions in the presence of perfect fluid which, as will be argued, should be
non-minimally coupled to the dS distance scale $\ell$. Despite the apparently non-minimal coupling, the theory is phenomenologically viable and does not lead to drastic violations of the equivalence principle. We clarify this in Section \ref{matter}, where it is shown that particles move along geodesics of an integrable Weyl geometry. For completeness we also consider a scalar field coupled to the dS gauge theory, in order to confirm the consistency of the cosmological set-up beyond the perfect fluid parameterisation. The conclusions are stated in Section \ref{conclu}. 
                

\section{The dS gauge theory}
\label{gauge}

The dS space is considered as the 4-dimensional quotient of the dS group by the Lorentz group. Consequently, in this Section
we will be referring to various distinct sets of coordinates. For clarity, the following table  
\begin{center}
\begin{tabular}{ c c c }
 coordinates & algebra & metric \\ 
 \hline
 $\{x^i\}_{i=1,2,3}$     & $\mathfrak{so}(3)$ & $\delta_{ij}=\text{diag}(1,1,1)$ \\ 
 $\{x^a\}_{a=0,1,2,3}$        & $\mathfrak{so}(3,1)$ & $\eta_{ab}=\text{diag}(-1,1,1,1)$ \\  
 $\{X^A\}_{A=0,\dots,4}$ & $ \mathfrak{so}(4,1)$ & $\eta_{AB}=\text{diag}(-1,1,1,1,1)$ \\  
 $\{x^\mu\}_{\mu=0,1,2,3}$ & ${\mathfrak{gl}_\parallel(3,1)}$ & $g_{\mu\nu}$ locally $\eta_{\mu\nu}$
\end{tabular}
\end{center}
summarises our conventions. 

The generators of the dS algebra $\mathfrak{so}(4,1)$ satisfy the commutation relations 
\be \label{comm}
[\Omega_{AB},\Omega_{CD}] = 2\lp \eta_{D[A}\Omega_{B]C} -  \eta_{C[A}\Omega_{B]D}\rp\,,
\ee
with $\eta_{AB}$ as given above. The 10 distinct generators $\Omega_{AB}=-\Omega_{BA}$ can be interpreted 
as spacetime rotations in 5 dimensions, whilst our spacetime has the 4-dimensional tangent space with the coordinates $\{x^a\}_{a=0,1,2,3}$ and the
metric $\eta_{ab}$. Therefore we consider the 4-dimensional rotations to the generated by
$\Omega_{ab}$ that coincide with the corresponding $\Omega_{AB}$, but define the rotations around the 5$^{\text{th}}$ dimension as
\be \label{pi}
\Pi_a = \ell^{-1}\Omega_{4a}\,.
\ee
The 4 generators $\Pi_a$ will be interpreted as (generalised) translations.
The algebra inherited from (\ref{comm}) by the new generators is
\begin{subequations}
\label{comm2}
\ba
\lb \Omega_{ab},\Omega_{cd}\rb & = & 2\lp \eta_{d[a}\Omega_{b]c} -  \eta_{c[a}\Omega_{b]d}\rp\,, \\
\lb \Pi_a,\Omega_{bc}\rb & = & 2\eta_{a[b}\Pi_{c]}\,, \label{comm2b}\\
\lb \Pi_a,\Pi_b \rb & = & -\ell^{-2}\Omega_{ab}\,. \label{comm2c}
\ea
\end{subequations}
The $\ell$ is a dimensionful parameter that quantifies how much boost along the 5$^{\text{th}}$ dimension is needed for a unit translation. In the 
limit $1/\ell \rightarrow 0$, (\ref{comm2}) reduces to the Poincar{\'e} algebra $\mathfrak{iso}(3,1)$ and the $\Pi_a$ become ordinary translations. 

In general, the form of $\Pi_a$ will depend upon the geometry of the symmetry breaking. 
To illustrate the embedding of the hyperboloid, let us consider here the flat slicing
\begin{subequations}
\label{coords}
\ba
X^0 & = & \ell\sinh{(t/\ell}) + e^{t/\ell}\delta_{ij} x^i x^j/2\ell\,, \\
X^i  & = & e^{t/\ell}x^i\,, \\
X^4 & = & \ell\cosh{(t/\ell}) - e^{t/\ell}\delta_{ij} x^i x^j /2\ell\,,
\ea
\end{subequations}
since then the induced metric $g_{ab}$ has the most commonly used cosmological (isotropic and homogeneous) form
\be
\eta_{AB}\diff X^A\diff X^B = g_{ab}\diff x^a\diff x^b = -\diff t^2 + e^{2t/\ell}\delta_{ij}\diff x^i\diff x^j\,. \nn
\ee
By inverting (\ref{coords}),
\begin{subequations}
\label{xa}
\ba
t\equiv x^0 & = &  \ell\log{(X^0+X^4)} - \ell\log{(\ell)}\,, \\
x^i & = & \ell X^i/(X^0+X^4)\,.
\ea
\end{subequations}
We find the conformal relations between the basis vectors,  
\begin{subequations}
\label{Xa}
\ba
\frac{\partial}{\partial X^a} & = & e^{-t/\ell}\partial_a\,, \\
\frac{\partial}{\partial X^4} & = & e^{-t/\ell}\lp \partial_t -\ell^{-1}x^i\partial_i\rp\,. 
\ea
\end{subequations}
Using the dictionary (\ref{coords},\ref{xa},\ref{Xa}) we can easily write down the 
orbital generators,
\begin{subequations}
\label{generators}
\ba
\Omega_{i0} & = & 2x_{[i}\partial_{0]} + (t+\ell\alpha)\partial_i\,, \\
\Omega_{ij} & = & 2x_{[i}\partial_{j]}\,, \\
\Pi_0 & = & \partial_t - \alpha\ell^{-1} x^i\partial_i\,, \\
\Pi_i & = & \beta\partial_i - \ell^{-1}x_i\lp \partial_t - \ell^{-1}x^k\partial_k\rp\,,
\ea
\end{subequations}
where we used the short-hands
\begin{subequations}
\label{ajab}
\ba
\alpha & \equiv & \frac{1}{2}\lp 1 + \ell^{-2}\delta_{jk}{x^j x^k} - e^{-2t/\ell}\rp = e^{-t/\ell}X^0/\ell\,, \nn \\
\beta & \equiv & \frac{1}{2}\lp 1 - \ell^{-2}\delta_{jk}{x^j x^k} + e^{-2t/\ell}\rp = e^{-t/\ell}X^4/\ell\,. \nn
\ea
\end{subequations}
As a cross-check, we verified that (\ref{comm2}) is satisfied by (\ref{generators}). 

An interesting conformal structure is manifest in the stereographic embedding \cite{Aldrovandi:2006vr} alternative to the flat slicing (\ref{coords}), though it
will be shown elsewhere that the Beltrami geometry \cite{Guo:2003qm} is convenient for the representations\footnote{In both pictures the $\Omega_{ab}$ have their usual form,
the ``transvection'' $\Pi_a$ becoming a translation contaminated with, in the stereographic coordinates the special conformal \cite{Pereira:2013zxa}, and in the Beltrami coordinates (as well as in the flat slicing) the ordinary conformal 
transformation.}.  

To gauge the $\mathfrak{so}(4,1)$, we now introduce the connection 1-form 
\be
\bA = \frac{1}{2}\bA^{AB}\Omega_{AB} =  \lp\frac{1}{2}\bA^{ab}{}_\mu\Omega_{ab} + \bA^a{}_\mu\Pi_a\rp\diff x^\mu\,. \label{afield}
\ee
The connection determines the dS-covariant (exterior) derivative $\Diff = \diff + \bA$, which further generates the field strength 2-form
$\Diff^2=\diff \bA + \bA\wedge\bA \equiv \bF$ and the 3-form identity $\Diff^3=\Diff\bF=0$. It is crucial to note
that because of the definition (\ref{pi}), we have $\bA^a = \ell\bA^{4a}$ and consequently \cite{Jennen:2014mba,Jennen:2015bxa,Jennen:2016pqw}
\be
\ell A^{a4}{}_{\mu,\alpha} =  A^a{}_{\mu,\alpha} - \log{\ell}_{,\alpha}A^a{}_\mu\,. 
\ee
As a result, the components of the field strength are slightly modified,
\begin{subequations}
\label{fs3}
\ba
F^a{}_{\mu\nu} & = & 2\lp A^a{}_{[\nu,\mu]} + A^a{}_{b[\mu}A^b{}_{\nu]}\rp - 2\log{\ell}_{,[\mu}A^a{}_{\nu]}   \,,  \,\,\, \\ \label{fs3}
F^{ab}{}_{\mu\nu} & = & 2\lp A^{ab}{}_{[\nu,\mu]} + A^a{}_{c[\mu}A^{cb}{}_{\nu]} - \ell^{-2} A^{[a}{}_\mu A^{b]}{}_\nu\rp\,.\,\,\,
\ea
\end{subequations}
Thus, a novel term that depends on the dynamics of the scale field $\ell$, now appears in the translation gauge field strength.
 
In addition to the connection 1-form (\ref{afield}), a symmetry-breaking scalar field $\xi^a$ is required. Otherwise one cannot introduce the coframe field $\bbe^a$, which is the 1-form defined by  
\be
\bbe^a = \bA^a +\Diff\xi^a\,. \label{tetrad}
\ee   
From this object, we further obtain the torsion 2-form $\bT^a=\Diff\bbe^a$ and the 3-form identity $\Diff\bT^a = \bF^a{}_b\wedge\bbe^b$. 
One sees that the translation gauge field strength coincides with the torsion 2-form once the Lorentz curvature $\bF^a{}_b=0$ is taken to vanish, since from (\ref{tetrad}) we have that 
$\bT^a = \Diff \bA^a + \bF^a{}_b \xi^b$. Assuming 
that the coframe field has an inverse, all the standard ingredients of gravitational geometry can now be constructed. In the language of Ref. \cite{Westman:2014yca}, our fundamental 
fields are $V^A$ and $\bA^{AB}$, and $\ell$ corresponds to the norm of the $V^A$ and the $\xi^a$ to the rest of its independent components, such that the definition (\ref{tetrad})
ensures the co-covariance of the components of $\bbe^a= \e^a{}_\mu\diff x^\mu$ and its inverse.  Then we can freely project the tangent space indices to 
spacetime indices and vice versa. 

In particular, we obtain the the spacetime torsion tensor $T^\alpha{}_{\mu\nu}$, and can then construct an action for a translation gauge theory 
in terms of the invariant $T$ known as the torsion scalar,
\be \label{torsionT}
T = \frac{1}{4}T^\alpha{}_{\mu\nu} T_{\alpha}{}^{\mu\nu} +
 \frac{1}{2}T^\alpha{}_{\mu\nu} T^{\nu\mu}{}_\alpha - T^\nu{}_{\mu\nu}T^{\alpha\mu}{}_\alpha\,. 
\ee
If we denote $\bar{\bT}^a$ the torsion 2-form in the limit when the evolution of dS scale is neglected, (\ref{fs3}) tells that $\bT^a=\bar{\bT}^a-\diff\log{\ell}\wedge\bA^a$.  
Plugging this into (\ref{torsionT}) we obtain that
\be
T = \bar{T} + 4\log{\ell}_{,\mu}\bar{T}^\mu - 6\lp\partial\log{\ell}\rp^2\,, \label{torsionT2}
\ee
where we defined $T_\mu = T^\alpha{}_{\mu\alpha}$. The equivalent result was reported in \cite{Jennen:2015bxa}. However, our action integral over this scalar,
\be \label{JPaction}
I_{\text{dS}} = -\frac{1}{2}\int \diff^4x \e \lb  \ell^{-2}\bar{T} + 4\ell^{-3}\ell_{,\mu} \bar{T}^\mu- 6\ell^{-4}\lp\partial{\ell}\rp^2\rb\,,  
\ee
has now different scalings for each of the terms, due to the dilatonic role of the dS scale. This action turns out to be equivalent to the conformally coupled
scalar-tensor theory \cite{Dirac:1973gk}. By recalling that the metric Ricci scalar ${R}$ is related to the torsion scalar via ${R} =-\bar{T}-2\ie\partial_\mu(\e \bar{T}^\alpha)$, we 
can rewrite (\ref{JPaction}) in the much more conventional (though pedantically speaking, ill-defined due to higher derivatives) scalar-tensor form
\be  \label{JPactionC}
I_{\text{dS}} =  \frac{1}{2}\int \diff^4x \e \lb  \ell^{-2}{R} +  6\ell^{-4}\lp\partial{\ell}\rp^2\rb\,.
\ee
It is well-known that this theory is invariant under the Weyl rescalings\footnote{A complete classification of scale invariance(s) in the general metric-affine geometry was given in Ref. \cite{Iosifidis:2018zwo}. Scale transformations in torsional geometry have been considered in e.g. \cite{Maluf:1985fj,Maluf:2011kf,Bamba:2013jqa,Wright:2016ayu,Lucat:2017wtu,Barnaveli:2018dxo,Formiga:2019frd}.} \cite{Dirac:1973gk}
\be \label{rescaling}
g_{\mu\nu} \rightarrow f^2 g_{\mu\nu}\,, \quad \ell \rightarrow f\ell\,.  
\ee
This scale invariance allows to reduce the theory explicitly to general relativity in the gauge $f=\ell_P/\ell$, but the symmetry (\ref{rescaling}) means that is an equivalence holds regardless of the gauge choice.

\subsection{On alternative formulations}
\label{alternatives}

It could also be interesting to reconsider the geometrical foundation \cite{BeltranJimenez:2019tjy} of the above formulation. In particular, since the model spaces are 
characterised by different scales, it may not be justified to consider 
the generators to be independent of the coordinates $x^\mu$. In particular, as we see in the cosmology-motivated example (\ref{generators}), the spacetime dependence enters into the
generators via the $\ell=\ell(x)$. The potential problem with this could however be avoided by reformulating the theory in a torsion-free geometry. 
Then one would to begin, instead of (\ref{pi}), with generators defined by the opposite scaling 
\begin{subequations}
\ba \label{pi2}
\hat{\Pi}_a & = & \Omega_{4a} = \hat{\eta}_{ab}\Pi^b\,,  \\
\hat{\Omega}_{ab} & = & \ell^{2} \Omega_{AB}\delta^A_a\delta^B_b = \hat{\eta}_{ac}\Omega^c{}_b\,. 
\ea
\end{subequations}
The same algebra (\ref{comm2}) in terms of the newly defined generators has to be then written in terms of the conformally rescaled metric $\hat{\eta}_{\mu\nu}=\ell^2\eta_{ab}$ 
as 
\begin{subequations}
\label{comm4}
\ba
\lb \hat{\Omega}_{ab},\hat{\Omega}_{cd}\rb & = & 2\lp \hat{\eta}_{d[a}\hat{\Omega}_{b]c} -  \hat{\eta}_{c[a}\hat{\Omega}_{b]d}\rp\,, \\
\lb \hat{\Pi}_a,\hat{\Omega}_{bc}\rb & = & 2\hat{\eta}_{a[b}\hat{\Pi}_{c]}\,, \label{comm3b}\\
\lb \hat{\Pi}_a,\hat{\Pi}_b \rb & = & -\ell^{-2}\hat{\Omega}_{ab}\,. \label{comm3c}
\ea
\end{subequations}
This indeed suggests a relation to the Weyl gauge theory and a rationale for the emergence of the scale symmetry (\ref{rescaling}). In this basis, the theory can be formulated consistently using the stereographic projection where the induced metric is conformally flat and the rotations are $\ell$-independent. In the end, the $\diff \log{\ell}$ term does not appear in the $\hat{\bT}^a$, but a corresponding term is found in the $\hat{\bF}^a{}_b$, and one can write down the usual quadratic curvature action that reduces to the dS general relativity.  We will not pursue here the details of this formulation (presumably resulting in the equivalent (\ref{JPactionC})). In fact, the geometry of the canonical version would be both torsion-free and flat \cite{Koivisto:2019jra,BeltranJimenez:2020sih}, but such a formulation would require the enlarging of the gauge group and be superfluous for the present purpose.   

\section{Cosmological solution}
\label{cosmo}

Cosmologies inspired by the Jennen-Pereira model \cite{Jennen:2015bxa} were analysed as a dynamical system by Otalora \cite{Otalora:2014aoa}. However, the class of models studied therein includes neither the particular case of (\ref{JPaction}) nor the version of Ref. \cite{Jennen:2015bxa} (obtained from \ref{torsionT2}), because, firstly, the sign of the scalar field kinetic term in Ref. \cite{Otalora:2014aoa} was flipped, and secondly, because the trace-coupling was considered as a function of the scalar field\footnote{On more general scalar-torsion modified gravity, see e.g. \cite{Bamba:2013jqa,Hohmann:2019gmt,Emtsova:2019qsl,Flathmann:2019khc,Golovnev:2018wbh,Raatikainen:2019qey,Bahamonde:2020cfv,Hohmann:2020dgy}. However, already linear perturbations \cite{Hohmann:2020vcv} indicate \cite{Golovnev:2018wbh,Raatikainen:2019qey} that generic such models are not viable. This stems from their Lorentz violation \cite{Li:2010cg}.}. We shall now explore the cosmology of the dS gauge theory (\ref{JPaction}), and find that is qualitatively different from models of scalar-torsion modified gravity.   

The line element in the flat Friedmann-Lema{\^i}tre-Robertson-Walker cosmology is
\be
\diff s^2 = -n^2(t) \diff t^2 + a^2(t)\delta_{ij}\diff x^i\diff x^j\,, 
\ee
where $n$ is the lapse function and $a$ the scale factor. In addition to these two metric components (of which $n$ can always be trivialised by simply a redefinition of $t$), we have the dS scale $\ell$ which may now evolve in time. We'll denote the expansion rates as follows:
\begin{center}
\begin{tabular}{ c c c }
 scale factor & variable & rate \\ 
 \hline
 temporal     & $n(t)$ & $N=\dot{n}/n$ \\ 
 spatial        & $a(t)$ & $H=\dot{a}/a$ \\  
 dimensional & $\ell(t)$ & $L=\dot{\ell}/\ell$    
\end{tabular}
\end{center}
Including a matter source $I_M$ describing a perfect fluid with the energy density $\rho_M$ and pressure $p_M$ coupled to the dS gravity (\ref{JPaction}), the cosmological mini-superspace action becomes
\be \label{action2}
I = I_{\text{dS}} + I_M = -\int \diff t \lb \frac{3a^3}{\ell^2 n}\lp H - L\rp^2 + na^3\rho_M\rb\,.  
\ee
The 1$^{\text{st}}$ and the 2$^{\text{nd}}$  Friedmann equations (obtained from the variations of $I$ wrt $n$ and $a$, respectively), can now be written as
\begin{subequations}
\label{friedmann}
\ba \label{friedmann1a}
3H^2 & = &  {\lp \ell n\rp^2}  \lp \rho_\ell + \rho_M\rp\,,  \\ 
\label{friedmann2a}
-2\dot{H}-3H^2 + 2NH & = & \lp \ell n\rp ^2\lp p_\ell + p_M\rp\,, 
\ea
\end{subequations}
where
\ba
\lp \ell n\rp^2 \rho_{\ell} &  = &  -3L^2 + 6HL\,, \nn  \\
\lp\ell n\rp^2  p_\ell & = & -2\dot{L} - 4HL + L^2 + 2NL\,. \nn
\ea
In vacuum, $\rho_M=0$, with the time slicing $n=1$, the solutions are $\ell/a=\text{constant}$. This already yields the insight into the theory that only the relative calibration of the two scale factors is fixed in vacuum, and neither of the scale factors alone. 

To properly couple matter sources to dS gravity with an evolving $\ell$, we should take into account the scaling of the energy density with $\ell$. 
For the purposes of background cosmology, the energy density of matter with an equation of state $w_M=p_M/\rho_M$ is then given by, up to a constant,
\be
\rho_M \sim a^{-3(1+w_M)}\ell^{-1+3w_M}\,. \label{prescription}
\ee  
This prescription results in the scaling one would expect from physical arguments in the cases of radiation or dust in the matter sector or, spatial curvature or a cosmological constant in the geometric sector. In particular, the effective action for a point particle, studied in more detail in Section \ref{particle}, suggests the scaling
$\rho_M \sim \ell^{-1}a^{-3}$ when $w_M=0$, and the scale invariance of the radiation is compatible with that $\rho_M \sim a^{-4}$, independently of $\ell$, when $w_M=1/3$. 
(Furthermore, the energy density due to a cosmological term is $\sim \ell^{-4}$ and independent of $a$, whilst the effective energy of a spatial curvature term $\sim (\ell a)^{-2}$.) 

Since, according to (\ref{prescription}), the field $\ell$ now couples non-minimally to matter, its equation of motion acquires a source term and reads
\ba
 \dot{H}  -   \dot{L} & + & \lp 2H-L - N\rp \lp H - L\rp  \nn \\
 & =  & \lp \frac{1}{6} - \frac{1}{2} w_M\rp \lp \ell n\rp ^2\rho_M \,. \label{nEoM}
\ea
It should be noted that in general, when $L \neq 0$, the energy densities obey the modified continuity equations, 
\begin{subequations}
\label{continuity}
\ba
\frac{\diff}{\diff t} \lp\ell^2 {\rho}_\ell\rp + 3H\ell^2\lp \rho_\ell + p_\ell\rp & = & -2L \ell^2 \rho_M\,,\,\, \label{continuitymod} \\ 
\dot{\rho}_M + 3H\lp 1+w_M\rp\rho_M & = & L \lp 1-3w_M \rp\rho_M\,.\,\, 
\ea
\end{subequations}
It is easy to see that the $\ell^2 = 8\pi G$, $L=0$ is a solution to the Friedmann equations (\ref{friedmann}).
Thus it is clear that the model defined above at least contains viable solutions that describe the standard cosmological background evolution. In the case of a possible time-evolution of $\ell$, more general solutions exist to the system of equations. 

To investigate such more general solutions, we begin with the power-law ansatz
\be
 n=1\,, \quad a \sim t^\alpha\,, \quad \ell \sim t^\lambda\,. 
\ee
By plugging this ansatz into the 1$^{\text{st}}$ Friedmann equation (\ref{friedmann1a}), we readily see that the power-laws must have the relation
\be
\lambda = \frac{1}{1+3w_M}\lb 3\lp1+w_M\rp\alpha -2\rb\,. \label{lambda}
\ee
The solution that gives back the expansion law of general relativity is $\lambda=0$ which implies $\alpha=2/(3+3w_M)$, but this is only one amongst the 1-parameter family of solutions parameterised by $\lambda$. Remarkably, these solutions satisfy also the 2$^{\text{nd}}$ Friedmann equation (\ref{friedmann2a}), and consequently they satisfy identically the equation of motion (\ref{nEoM}) as well.  Accelerating solutions exist. For a background fluid with $w_M>-1/3$, the universe accelerates as if dominated by a quintessence-like field given that $\lambda>-1$, and further, the universe super-accelerates if $\ell<-2/(1+3w_M)$. In general, the universe expands as if was filled with a fluid that has the equation of state
\be \label{weff}
w = \frac{\rho_\ell + \rho_M}{p_\ell + p_M} = \frac{w_M-\lp \frac{1}{3} + w_M\rp\lambda}{1+\lp \frac{1}{3} +w_M\rp\lambda}\,.
\ee  
More general cosmological solutions, sourced by perfect fluids with $\dot{w}_M \neq 0$ (which can also effectively describe several distinct perfect fluid components), could be studied numerically.

It is worthy to point out that the dS coupling prescription (\ref{prescription}) is essentially the unique viable possibility. We will briefly comment upon some alternative prescriptions, omitting the details of the derivations. The Jennen-Pereira model  \cite{Jennen:2015bxa} with the standard coupling prescription (i.e. $\rho_M \sim a^{-3(1+w_M)}$) is not compatible with cosmological evolution\footnote{More precisely, the Friedmann equations would be consistent only for stiff fluid matter $w_M=1$. This was first pointed out to us by Sergio Bravo Medina.}. If this model is supplemented with the $\ell$-dependent cosmological constant (the sign has to be negative), cosmological evolution can be recovered such that in the standard Friedmann equation $G \rightarrow G/(1+3w_M)$, and therefore in the radiation dominated era the effective gravitational coupling would be $G/2$, that appears too drastical modification to allow viable early universe phenomena such as nucleosynthesis and the formation of the cosmic microwave background. 
Yet, one could further adjust the model by retaining the minimal matter coupling but taking into account the $\ell$-dependence of the gravitational coupling. In such a prescription a radiation-dominated era is not only phenomenologically excluded, but incompatible with the Friedmann equations in the first place. 

Thus, it turns out that the dS matter coupling prescription (\ref{prescription}) that we justified by physical principles, could actually have been formally deduced by requiring the existence of viable cosmological background solutions.  

\subsection{On the relevance of the solution}

It is lluminating to show that the family of solutions is indeed equivalent under the symmetry (\ref{rescaling}). The invariant combination of the metric and the
scale field is $\hat{g}_{\mu\nu} = (\ell_P/\ell)^2g_{\mu\nu}$, and correspondingly we denote $\hat{a}=(\ell_P/\ell) a$, and $\hat{t}$ the time coordinate when the lapse function is $\hat{n}=(\ell_P/\ell)$. It is then straightforward to compute
the invariant Hubble rate and its time derivative,
\begin{subequations}
\ba
\hat{H} & = &  (\ell/\ell_P)\lp H- L\rp\,,  \\ 
\frac{\diff}{\diff \hat{t}}\hat{H} & = &   (\ell/\ell_P)^2\lb \dot{H}- \dot{L} - (H-L) L \rb\,. 
\ea
\end{subequations}
Plugging in the relations (\ref{lambda}) and (\ref{weff}) we obtain the result for the expansion rate that corresponds to the effective equation of state 
$\hat{w} = w_M$. In terms of the  ``hatted'' variables, the Friedmann equations (\ref{friedmann}) of course assume their standard form. 

The gauge freedom allows a radically different re-interpretation of the expanding universe. In an extreme case, we can understand all the observational data in a static universe (obtained by setting $\alpha=0$
in the above family of solutions), where instead the gravitational coupling as well as the masses of particles are evolving in time (according to $\lambda=-2/(1+3w_M)$). For example, the observed
cosmological redshift  of photons is then not due to the stretching of the wavelengths together with the spatial scale factor $a(t)$, but it is due to the shrinking of the dimensional scale factor $\lambda(t)$. 
In this description of the universe, we clearly have no curvature singularity and therefore the cosmological spacetime appears to be non-singular and extendable to $t \rightarrow -\infty$. Going backward in time from the present, the dS scale grows indefinitely, and the big bang would-be-singularity occurs at the point wherein the dS scale becomes infinite and the hyperboloid flattens out (this is the contraction limit $\mathfrak{so}(4,1) \rightarrow \mathfrak{iso}(3,1)$), and continuing this naive extrapolation to still earlier times, the geometry becomes that of anti-dS with the radius now 
shrinking indefinitely as we wind backwards towards $t \rightarrow -\infty$. In this frame, both the metric and the total curvature invariants are identically zero, though the torsion scalar (\ref{torsionT}) is $T=6L^2$ and thus diverges at $t=0$. The physical matter quantities remain finite. The radiation\footnote{If dust is present at such a primordial stage, its energy density momentarily disappears at $t=0$. This might be relevant in regards the initial conditions for the geometric dark matter discovered in \cite{Zlosnik:2018qvg}. We note that at least the naive prescription (\ref{prescription}) excludes sources with $w_M>1/3$, since their energy density would diverge.} energy density and the pressure are always constant in this static universe frame, as seen from (\ref{prescription}).  
 
The possible relevance of the dS kinematics to a new cosmological paradigm has been foreseen in some discussions \cite{Aldrovandi:2004km,Araujo:2015oqa}. Though no solutions were presented, and the focus was on the opposite contraction limit $\ell \rightarrow \infty$, the main insight that the conformal property of the (apparently singular) transition point could be the key in connecting two aeons in sir Penrose's conformal cyclic cosmology  \cite{Araujo:2015oqa}, is strongly corroborated by our exact cosmological solution in the consistent dS gauge gravity (\ref{JPaction}). By adopting Willem de Sitter's own, projective view of the dS geometry \cite{McInnes:2003xm,Ong:2016vwr}, the solution might naturally be enclosed into the eternal return of the aeon that is our unique universe.  
 
In the more mainstream context of string theory, the existence of negative energy vacua seems to be not only a generic prediction in the landscape of myriad universes, but a requirement for the consistent definition of an S-matrix, and it has proven quite a challenge to find ways that may lead to positive vacuum energies compatible with the one observed universe \cite{Kachru:2003sx,Dasgupta:2019gcd}. Previous attempts at realising a non-singular anti-dS to dS transition have resorted to rather complicated mechanisms requiring various new indgredients for their realisation \cite{Biswas:2011qe,Gupt:2013poa}. It is remarkable that we seem to consistently predict the desired non-singular transition, based on an action that is locally equivalent to general relativity, but underpinned by the principles of dS gauge theory. -It should be noted that the reinterpretation of the cosmological expansion as a variation of mass scales is of course well-known in the context of Fierz-Jordan-Brans-Dicke theory, and in particular, the possibility that the big bang singularity is a field coordinate singularity (i.e. removable by a change of variables) was introduced and clarified by Wetterich \cite{Wetterich:2013jsa,Wetterich:2013aca,Wetterich:2020oyy} (in the context of his theory of Variable Gravity \cite{Wetterich:2013jsa}, which is not a mere reformulation of general relativity). The removal of black hole singularities was also considered, employing more general than conformal change of field coordinates \cite{Domenech:2019syf}.

Finally, let us mention that Hohmann {\it et al} \cite{Hohmann:2018shl} have recently pointed out that the theory (\ref{JPactionC}) formally contains vacuum solutions with wormholes, despite their local equivalence with the standard vacuum solutions. The physicality of such wormholes hinges on global, topological issues. As Hohmann {\it et al} \cite{Hohmann:2018shl} explained, the key point is that the solutions may be related by improper Weyl transformations (\ref{rescaling}), where the factor $f$ may vanish or become infinite at some points (in other words, the Jacobian of the field coordinate transformation is not defined at those points). It is precisely in this sense that the dS gauge theory (\ref{JPaction}) is inequivalent to general relativity, and can thus accommodate a more general variety of physically distinct solutions. 
 
\section{Implications to matter}
\label{matter}

At the level of background cosmology it was sufficient to exploit the perfect fluid parameterisation (\ref{prescription}) for matter sources, but the question may remain whether the proposed dS coupling prescription is consistent for more fundamental field theory description of massive matter fields. To address this question, we consider the action for a point particle  and for a scalar field. 

\subsection{Point particle}
\label{particle}

Consider the massive point particle action,
\be
I_{pp} = \int m\diff s\,, 
\ee
where the line element for a time-like curve $x^\mu$ is $\diff s = \sqrt{-g_{\mu\nu}\diff x^\mu\diff x^\nu}$ and the
mass $m$ is related to the fundamental scale $\ell$ as $m(x)=m_0/\ell(x)$, $m_0$ being the dimensionless constant of proportionality. 
We consider small variations $\delta x^\mu$ of the curve $x^\mu(\tau)$ parameterised by an arbitrary parameter $\tau$, 
\be
\delta I_{pp} = \int \lb \delta m \frac{\diff s}{\diff\tau} - \frac{m}{2\diff s}\delta\lp g_{\mu\nu}\dot{x}^\mu\dot{x}^\nu\rp\rb\diff\tau \,.  
\ee
The first term we can write as $\int \delta m {\diff s} = \int m_{,\mu} \dot{x}^\mu\diff s$ and for the second term we apply integration by parts.
Adding the two terms we get
\ba
\delta I_{pp} & = &  \int\Big[ m g_{\mu\nu} \frac{\diff^2 x^\nu}{\diff s^2}  +  m \frac{\diff x^\alpha}{2\diff s}\frac{\diff x^\nu}{\diff s}\lp 2 g_{\mu(\nu,\alpha)} - g_{\alpha\nu,\mu}\rp  \nn \\ \nn
& + & \frac{\diff x^\alpha}{2\diff s}\frac{\diff x^\nu}{\diff s}\lp 2 g_{\mu(\nu}m_{,\alpha)} - g_{\alpha\nu}m_{,\mu}\rp \Big]\delta x^\mu\diff s =0\,.
\ea
Since this holds for arbitrary variations of the path $\delta x^\mu$, we get, by raising one index and dividing by $m$,
\ba
\frac{\diff^2 x^\alpha}{\diff s^2}  & + & \frac{1}{2}g^{\alpha\beta}\lp g_{\beta\mu,\nu} + g_{\beta\nu,\mu} - g_{\mu\beta,\mu}\rp \frac{\diff x^\mu}{\diff s}\frac{\diff x^\nu}{\diff s}\nn \\
= & - &  \frac{1}{2m}\lp \delta^\alpha_\mu m_{,\nu} + \delta^\alpha_\nu m_{,\mu} - g_{\mu\nu}{m}^{,\alpha} \rp \frac{\diff x^\mu}{\diff s}\frac{\diff x^\nu}{\diff s}\,.\nn
\ea 
Since $\log{m}_{,\alpha} = -\log{\ell}_{,\alpha}$, this can be written as
\be \label{massivegeodesic}
\ddot{x}^\alpha + \overset{\star}{\Gamma}{}^\alpha{}_{\mu\nu}\dot{x}^\mu \dot{x}^\mu = 0\,, 
\ee
where the overdot now denotes the derivative wrt the proper time $\tau=s$ and the connection is
\be \label{connectiondecomp1} 
\overset{\star}{\Gamma}{}^\alpha{}_{\mu\nu} = \left\{^{\phantom{i} \alpha}_{\mu\nu}\right\} - \lp \delta^\alpha_{(\mu}\log{\ell}_{,\nu)}- \frac{1}{2}g_{\mu\nu}\log{\ell}^{,\alpha}\rp\,.
\ee
Thus, we predict that the matter moves along the geodesics of a Weyl connection, for which the Weyl gauge field $\ell_\mu=\ell_{,\mu}$ is pure gauge and thus its curvature vanishes $F_{\mu\nu} = 2\ell_{[\mu,\nu]}=2\ell_{,[\mu\nu]}=0$. Therefore there is no second clock effect. For more details and the extension of the integrable Weyl (sometimes called semi-metric) geometry to generic non-metric geometry, see \cite{BeltranJimenez:2020sih}.  

In terms of the arbitrary parameter $\tau$, (\ref{massivegeodesic}) generalises to
\be \label{massivegeodesic2}
\ddot{x}^\alpha + \overset{\star}{\Gamma}{}^\alpha{}_{\mu\nu}\dot{x}^\mu \dot{x}^\mu = -\dot{s}^2\frac{\diff^2 \tau}{\diff s^2}\dot{x}^\alpha\,. 
\ee
We see that the reparameterisation $\tau = a s + b$ where $a,b$ are constants, does not change the form of the equation (\ref{massivegeodesic}). We also note that the
projective transformation of the connection by a one-form $p_\mu$ can be compensated by the reparameterisation that satisfies
\be
\overset{\star}{\Gamma}{}^\alpha{}_{\mu\nu} \rightarrow \overset{\star}{\Gamma}{}^\alpha{}_{\mu\nu} +\delta^\alpha_\nu p_\mu\,, \quad 
p_\mu\dot{x}^\mu = \dot{s}^2\frac{\diff^2 \tau}{\diff s^2}\,.
\ee
A reparameterisation of the curve thus corresponds to a projective transformation of the affine geometry. The curve abstracted from its parameterisation i.e. the projective equivalence class of the geodesic is called a path. 

A more elementary modelling of matter fields would begin with spinor fields, but in the end the classical approximation relevant to our purposes is given by the point particle action where the $m \sim \ell^{-1}$ is inherited from the spinor mass term. Spinor fields and gauge fields can be coupled to dS gravity elegantly with polynomial Lagrangians \cite{Pagels:1983pq,Westman:2012zk}. 

\subsection{Scalar field}

Considering a self-interacting scalar field, our coupling prescription suggests the Lagrangian
\be
I_\phi = -\int \diff^4 x\sqrt{-g} \lb \frac{1}{2\ell^2}\lp \partial\phi\rp^2 + \ell^{-4}V(\phi)\rb\,. 
\ee
In the cosmological setting, this gives the total action $I=I_{\text{dS}}+I_\phi$ as
\be
I=-\int\diff t a^3\lb \frac{3}{\ell^2 n}\lp H-L\rp^2 - \frac{1}{2\ell^2}\frac{\dot{\phi}^2}{n} + \frac{n}{\ell^4}{V(\phi)} \rb\,.
\ee
The scalar field contribution to the Friedmann equations is then given by
\begin{subequations}
\ba
\rho_\phi & = & \frac{1}{2}\lp{\dot{\phi}}/{\ell n}\rp^2 + {\ell^{-4}}{V(\phi)}\,,  \\
p_\phi & = & \frac{1}{2}\lp{\dot{\phi}}/{\ell n}\rp^2 - {\ell^{-4}}{V(\phi)}\,. 
\ea
\end{subequations}
The Klein-Gordon equation is obtained by the variation wrt the scalar field,
\be \nn
\ddot{\phi} + \lp 3H - 2L-N\rp\dot{\phi} + n^2 \ell^{-6}V'(\phi)=0\,,
\ee
This equation, by multiplying with $\dot{\phi}$ and rearranging the terms, reduces to (\ref{continuitymod}). 

To verify the consistency of the system, we consider also the equation of motion 
\ba
 \dot{H}  -   \dot{L} & + & \lp 2H-L - \frac{\dot{N}}{N}\rp \lp H - L\rp   \nn \\
  & =  & \frac{1}{6}\lp -\dot{\phi}^2 + 4N^2\ell^{-2}V(\phi)\rp \nn \\ 
 & =  & \lp \frac{1}{6} - \frac{1}{2} w_\phi\rp \lp \ell N\rp ^2\rho_\phi \,,  \nn 
\ea
which is in full agreement with (\ref{nEoM}). Therefore this equation is, as it should, degenerate with the Friedmann equations. 

We note that even though a massless scalar field $V(\phi) \approx 0$ 
is a perfect fluid with the stiff fluid equation of state $w_\phi=1$, the heuristic perfect fluid coupling prescription (\ref{prescription}), which would naively suggest the kinetic term to scale as $\ell^2$, is not the correct one to adopt for a proper field 
theoretical description. The quadratic kinetic term $\sim (\partial\phi)^2$ inherits the scaling dimension $\sim [\phi^2] \sim [\ell^{-2}]$ from the dimension of the scalar field, and 
this is the fundamental rationale that determines the coupling.

\section{Discussion}
\label{conclu}

Motivated by the fact that our universe has fundamental limiting scales both in the infrared and in the ultraviolet ends of the spectrum, we developed a dS gauge theory of gravity which incorporates the new kinematic invariant $\ell$, besides the invariant $c$ of Einstein's relativity. The Cartan-geometric construction, illustrated upon the dS hyperboloid embedded into a spacetime with one extra dimension, was based upon nothing but the standard gauge field, a connection 1-form and a symmetry-breaking scalar field. Gravity was realised as a gauge theory of translations in the sense
that the action is quadratic in the translation gauge field strength $\bF^a$ whilst the homogeneous model spaces are flat $\bF^a{}_b=0$  (though, as mentioned in \ref{alternatives}, it would be possible to formulate a more canonical version of translation gauge theory). 

A key insight was that the theory (\ref{JPaction}) exhibits the rescaling invariance (\ref{rescaling}). Thus, the calibration of the scale $\ell$ is arbitrary, and it can changed without affecting the physics given the accompanying rescaling of the metric. Incidentally, the theory thus realises the foundational motivations of both the dS and the Weyl gauge theories. On one hand, the description of our universe requires observer-independent scales. On the other hand, absolute scales are physically meaningless. Thus, the new dS theory may provide a conceptually improved framework to the century-old problem of introducing scales into physics. From a formal point of view, the orthogonal symmetry is considerably neater than the Weyl extension of the Poincar\'e symmetry. An obvious direction to pursue in the future is the incorporation of the two limiting scales, $\ell_P$ and $\ell_\Lambda$ independently, via the completion of the dS to the conformal symmetry\footnote{It is natural to speculate that Dirac's original motivation for (\ref{JPactionC}), the large number hypothesis \cite{Dirac:1973gk,Ray:2007cc}, could be vindicated by exploiting the additional freedom provided by another scalar field, thus yielding the satisfactory explanation of various other scales in physics.}.

Let us comment on our theory also in view of the so-called ``teleparallel''  models of gravity. Modifications of gravity in that context are by now well-known to violate Lorentz symmetry, resulting in extra degrees of freedom which typically have strong coupling and other unwanted problems\footnote{Such concerns have been raised earlier in the literature \cite{Kopczynski_1982,Li:2010cg}, and the current state of art in the problematics of the extra degrees of freedom is reviewed in \cite{Blixt:2020ekl,Golovnev:2020zpv}.}. In contrast, the new theory developed in this article, though formulated in terms of a flat  connection $\bA^a{}_b$ i.e. in a ``teleparallel''  geometry, is not based on a violation but on an extension of the Lorentz symmetry. Thus our approach also seems to suggest a way of generating viable ``teleparallel''  gravity models. However, since the theory (\ref{JPaction}) we arrived at has the metric scalar-tensor equivalent (\ref{JPactionC}), it still remains an open question whether a consistent ``genuinely teleparallel'' modification of gravity is possible.   

The aim of this article was to explore the cosmology of the dS gauge theory with time-evolving distance scale $\ell=\ell(t)$. We derived the family of exact solutions which is characterised by the effective equation of state (\ref{weff}) for the gravity-fluid system. Incidentally, it turned out that the coupling prescription (\ref{prescription}) that follows from the physical interpretation of the dS scale $\ell$, is actually necessary for the existence of realistic cosmological solutions (including even those which reduce to the standard solutions in general relativity). We considered the possible reinterpretation of observations in the frame where the universe does not expand but the dS scale is evolving in time. This is, to our knowledge (despite the often-made claims otherwise in the vast literature on ``teleparallel'' cosmology), the first description of the cosmic geometry {\it de facto} in terms torsion, without the metric curvature playing its usual role. We argued that such a novel description could allow the consistent extension of cosmology beyond the big bang, and believe that the theory and its cosmological implications merit further investigation. 



\acknowledgements

TK would like to thank Manuel Hohmann and Hardi Veerm\"ae for insightful discussions on scale invariance, and Sergio Bravo Medina and David Mota for an earlier collaboration on a related topic. This work was supported by the Estonian Research Council grants PRG356 ``Gauge Gravity'' and MOBTT86, and by the European Regional Development Fund CoE program TK133 ``The Dark Side of the Universe''.


\bibliography{dSrefs}

\end{document}